\documentclass[12pt,english,aps,manuscript]{revtex4}
\usepackage{amsmath}

\input{tcilatex}

\begin{document}

\title{Bistable and dynamic states of parametrically excited ultrasound in a
fluid-filled cavity.}
\author{Isabel P\'{e}rez-Arjona, V\'{\i}ctor J. S\'{a}nchez-Morcillo and V%
\'{\i}ctor Espinosa}
\affiliation{Instituto de Investigaci\'{o}n para la Gesti\'{o}n Integrada de las Zonas
Costeras, Universidad Polit\'{e}cnica de Valencia, Crta. Natzaret-Oliva s/n,
46730 Grau de Gandia (Valencia), Spain}

\begin{abstract}
In this paper we have considered the problem of parametric sound generation
in an acoustic resonator filled with a fluid, taking explicitely into
account the influence of the nonlinearly generated second harmonic. A simple
model is presented, and its stationary solutions obtained. The main feature
of these solutions is the appearance of bistable states of the fundamental
field resulting from the coupling to the second harmonic. An experimental
setup was designed to check the predictions of the theory. The results are
consistent with the predicted values for the mode amplitudes and parametric
thresholds. At higher driving values a self-modulation of the amplitudes is
observed. We identify this phenomenon with a secondary instability
previously reported in the frame of the theoretical model. 
\end{abstract}

\maketitle

\section{Introduction}

Ultrasonic resonators are devices that confine the acoustic fields in a
finite region of the space. When driven by an external energy source (e.g.
vibrations of one of their boundaries), the reflection and resonance
conditions imposed by the boundaries provide the conditions to obtain high
amplitude fields, favoring the development of nonlinear effects, as the
emergence of frequencies different from that of the driving.

The knowledge of the field evolution in resonators in nonlinear regime is
important both in its fundamental and applied aspects. One of the simplest
configurations consists in two plane and parallel walls (an acoustic
interferometer) where nearly one-dimensional standing waves are formed along
the cavity axis. The system was the basis, forty years ago, of the first
experimental observation of acoustic parametric excitation in a fluid \cite%
{Korpel65,Adler68}. The phenomenon, first observed by Faraday \cite{Faraday}
and later described by Lord Rayleigh, consists in the emergence of
oscillation modes at frequencies smaller than that of the driving, when a
parameter of the system is varied periodically in time. The phenomenon is
universal, and has been demonstrated in a variety of physical systems \cite%
{Adler71}. In the case of the acoustical interferometer, the length of the
cavity (and thus the eigenfrequencies of the normal modes) is the
time-dependent parameter, and the parametric excitation is achieved when the
input energy is high enough to overcome the dissipative losses. The
parametric fields usually appear as doublets, whose frequencies $f_{1}$ and $%
f_{2}$ add to match the driving frequency $f_{0}$, i.e. $f_{0}=f_{1}+f_{2}$,
although in some circumstances the half-frequency mode is observed.

A theoretical description, based on the Mathieu equation, has been
successfully applied to the description of these processes in an acoustical
interferometer, allowing to predict the subharmonic spectrum and its
excitation threshold \cite{Adler71,Teklu07}. This approach has however a
restricted validity, since it can not describe the further evolution of the
parametric fields above the threshold and, on the other hand, it ignores the
unavoidable effects of the higher harmonics of the driving, which are the
main signature of non- or weakly-dispersive acoustic systems.

The spectrum of higher harmonics can be controlled using additional
dispersion mechanisms, such as bubbly fluids or walls with selective
(frequency dependent) absorption \cite{Zarembo75,Yen75}. Ostrovsky et al 
\cite{Ostrovsky78} designed a waveguide cavity where dispersion is
introduced by the lateral boundaries, and the higher harmonics could be
completely inhibited. As a theoretical framework, a set of evolution
equations for the trial of interacting modes was considered, resulting in an
excellent agreement with the measured amplitudes above threshold.

This theory has been also applied to the interferometer case \cite%
{Yen75,SanchezMor04,PerezArj05,SanchezMor06}, but the agreement with the
experiment \cite{Yen75} was mainly qualitative. The discrepancies can be
interpreted in terms of the influence of the first higher harmonics (those
with higher amplitudes) on the parametric process, introducing additional
features that can not be captured by the fully dispersive model.

In this work we present a theoretical description of parametric sound
generation in a fluid-filled interferometer, extending the previous models
to include the coupling with the second harmonic. The stationary solutions
both above and below the threshold are obtained and discussed. The solutions
are compared with experimentally measured values, with an excellent
agreement. The novel effects induced by the second harmonic (e.g.
bistability or hysteresis) are discussed. Finally, we also report the
existence of low frequency self-oscillations, occurring at pump values
beyond the parametric instability threshold. This phenomenon is discussed in
terms of secondary instabilities of the proposed model.

\section{Theory}

The acoustical interferometer considered in this paper is composed by two
parallel and solid walls, with thicknesses \textit{D}, located at a distance 
\textit{L} from each other, containing a fluid medium inside. Each medium
involved in the model is acoustically characterized by its density $\rho $,
bulk modulus $\kappa $, and sound velocity \textit{c}, related as $c^{2}=%
\frac{\kappa }{\rho }$. The resonance modes (eigenfrequencies) of the
resonator depend on these parameters, defining the acoustical impedances $%
z=\rho c$ of each section. In the ideal (loseless) case, corresponding to an
infinite impedance of the walls, the resonance modes are equidistant and
obey the equation%
\begin{equation}
\tan \left( k_{f}L\right) =0  \label{modes1}
\end{equation}%
where $k_{f}=2\pi f/c_{f}$. However, in a real system the impedances have
finite values, and the spectrum of the resonator is no longer equidistant,
but distributed according to the transcendental equation%
\begin{equation}
\tan \left( k_{f}L\right) =\frac{2\mathcal{R}\tan \left( k_{w}H\right) }{%
\mathcal{R}^{2}\tan ^{2}\left( k_{w}H\right) -1},  \label{modes2}
\end{equation}%
where $k_{w}=2\pi f/c_{w}$and $\mathcal{R}$ is the quotient between wall and
fluid impedances.

Now consider the driven system, assuming that one of the walls vibrate with
frequency $f_{0}$. Then, above the parametric generation threshold the
spectrum inside the resonator can be decomposed in two sets: the
subharmonics resulting from the parametric instability and the higher
harmonics of the driving, $nf_{0}$, with \textit{n} an integer number,
resulting from the weak dispersion in the system. The amplitude of any of
these harmonics decrease with the detuning, defined as the difference
between the corresponding field frequency $\omega _{n}$ and the frequency of
the closest cavity mode, $\omega _{n}^{c}$, \textit{i.e.} $\delta
_{n}=\omega _{n}-\omega _{n}^{c}$. The energy transfer into a harmonic is
then more effective under resonance conditions. In an ideal cavity, a
resonant driving implies that all higher harmonics are also resonant with a
cavity mode, enhancing the cascade of energy from the driving into many
higher frequency components. However, an initial detuning $\delta _{0}$
implies that the detuning of the higher order modes increases linearly with
the frequency (Fig. 1(a)). As a consequence, the amplitude of these modes is
reduced with respect to the resonant case. This situation can be more
dramatic in a real system, where the mode distribution is non-equidistant
and the detuning of the second and higher harmonics can be tuned to be much
larger than that of the driving and the subharmonics, as shown in (Fig. 1
(b)).

The above arguments suggest that a theoretical approach to parametric sound
generation must take into account the effect of the second harmonic, whose
amplitude is not negligible in general, but one can neglect the influence of
the third and higher harmonics, assuming that they are sufficiently out of
resonance. To simplify the analysis, we consider that the parametric
generation is degenerated, \textit{i.e.}, the half harmonic of the driving
is excited.

Under these assumptions, and following the technique described in detail in 
\cite{PerezArj05} for the fully dispersive case (neglecting higher
harmonics), the following system of equations for the evolution of the
slowly-varying amplitudes for each mode is obtained:%
\begin{align}
\frac{dP_{0}}{dt}& =-(\gamma _{0}+i\delta _{0})P_{0}-i\beta \left(
P_{1}^{2}+P_{2}P_{0}^{\ast }\right) +\frac{c}{L}P_{in},  \notag \\
\frac{dP_{1}}{dt}& =-(\gamma _{1}+i\delta _{1})P_{1}-i\frac{\beta }{2}%
P_{0}P_{1}^{\ast },  \label{fullmodel} \\
\frac{dP_{2}}{dt}& =-(\gamma _{2}+i\delta _{2})P_{2}-i\beta P_{0}^{2}, 
\notag
\end{align}%
corresponding to the fundamental (driving), subharmonic and second harmonic
respectively. Other parameters are: $P_{in}$ the driving amplitude (pump), $%
\gamma _{i}$ the decay rates of each mode in the cavity, $\beta =\frac{%
\omega _{0}}{4\rho c^{2}}\left( 1+B/A\right) $ is related to the
nonlinearity parameter $B/A$ of the fluid and $\delta _{i}=\omega
_{i}-\omega _{i}^{c}$ are the detunings.

The dynamical system given by Eqs.(\ref{fullmodel}) can be reduced to a
simpler, dimensionless form, defining the new variables $A_{0}=i\frac{%
2\gamma _{1}}{\beta }p_{0},A_{1}=i\frac{\sqrt{2\gamma _{0}\gamma _{1}}}{%
\beta }p_{1},A_{2}=i\frac{\gamma _{0}}{\beta }p_{2}$, $E=i\frac{2L\gamma
_{0}\gamma _{1}}{\beta }p_{in}$, and the parameters $\gamma =\frac{\gamma
_{1}^{2}}{\gamma _{0}\gamma _{2}}$ and $\Delta =\delta _{i}/\gamma _{i}$.
With these changes we obtain%
\begin{align}
\gamma _{0}^{-1}\frac{dA_{0}}{dt}& =-\left( 1+i{{\Delta }_{0}}\right) {A_{0}}%
+E-{A_{1}}^{2}-{A_{2}A_{0}^{\ast }},  \notag \\
\gamma _{1}^{-1}\frac{dA_{1}}{dt}& =-\left( 1+i{{\Delta }_{1}}\right) {A_{1}}%
+{A_{0}}\,{{A}_{1}^{\ast }},  \label{adimmodel} \\
\gamma _{2}^{-1}\frac{dA_{2}}{dt}& =-\left( 1+i{{\Delta }_{2}}\right) {A_{2}}%
+i\gamma {A_{0}^{2}}.  \notag
\end{align}

Equations (\ref{adimmodel}) admit two different stationary solutions. When
the pump amplitude \textit{E} is below the parametric threshold, one can set 
$A_{1}=0$. Neglecting the temporal derivatives in Eqs.(\ref{adimmodel}) we
obtain%
\begin{align}
E^{2}& =\left\vert A_{0}\right\vert \left( 1+{\Delta _{0}^{2}}+\frac{\gamma
\left\vert A_{0}\right\vert ^{2}\left( \gamma \left\vert A_{0}\right\vert
^{2}+2(1-{{\Delta }_{0}{\Delta }_{2}}\right) }{1+{\Delta _{2}^{2}}}\right)
^{1/2}  \label{fund} \\
\left\vert A_{2}\right\vert ^{2}& =\frac{\gamma ^{2}}{1+\Delta _{2}^{2}}%
\left\vert A_{0}\right\vert ^{4}  \notag
\end{align}%
where we have defined $I_{0}=\left\vert A_{0}\right\vert ^{2}$and $%
I_{2}=\left\vert A_{2}\right\vert ^{2}$.

The solution given in Eq.(\ref{fund}) reflects the fact that the amplitudes
of both the fundamental and the second harmonic grow with the pump
amplitude, as expected. The most interesting feature is however the
emergence of multivalued solutions, \textit{i.e.} the system shows a
bistable or hysteretic behavior, when some conditions on the parameters are
fulfilled. The condition for multivaluedness is found by imposing the
existence of an inflection point in the S-shaped curve given by Eqs.(\ref%
{fund}). This occurs when $\tfrac{d^{2}E^{2}}{dI_{0}}=0$, i.e. at%
\begin{equation}
\left\vert A_{0}\right\vert ^{2}=-\frac{2(1-\Delta _{0}\Delta _{2})}{3\gamma 
}.  \label{bistable}
\end{equation}

Note that the existence of bistable solutions requires that $\Delta
_{0}\Delta _{2}>1$, \textit{i.e. }both detunings must have the same sign and
exceed a certain critical value. Figure 2 shows the monostable (a), critical
(b) and bistable (c) cases of the fundamental mode, for different sets of
parameters.

At a pump value given by%
\begin{equation}
{E_{th}}=\sqrt{\left( 1+{\Delta _{1}^{2}}\right) \left[ \left( 1+\tilde{%
\gamma}\right) ^{2}+\left( \Delta _{0}-\tilde{\gamma}\Delta _{2}\right) ^{2}%
\right] }  \label{1stthreshold}
\end{equation}%
the solution given by Eq.(\ref{fund}) becomes unstable, denoting the
threshold of the parametric instability. The new solution corresponds to a
non-zero value of the subharmonic amplitude, given by 
\begin{equation}
\left\vert A_{1}\right\vert ^{2}=\left( -1+{{\Delta }_{0}{\Delta }}%
_{1}\right) -\tilde{\gamma}\left( 1+{\Delta }_{1}{{\Delta }_{2}}\right) \pm 
\sqrt{P^{2}-\left[ {\Delta }_{0}+{\Delta }_{1}+\tilde{\gamma}\left( {\Delta }%
_{1}-{\Delta }_{2}\right) \right] ^{2}}  \label{subharmonic}
\end{equation}%
where we have defined $\tilde{\gamma}=\gamma \frac{1+{\Delta _{1}^{2}}}{1+{%
\Delta _{2}^{2}}}$.

Above the instability threshold, the fundamental mode saturates to a
constant value%
\begin{equation}
\left\vert A_{0}\right\vert ^{2}=1+{\Delta }_{1}^{2},  \label{saturatefund}
\end{equation}%
and $A_{2}$ is given by Eq.(\ref{fund}).

Note that in the limiting cases of large second harmonic detuning $\Delta
_{2}$ or large losses $\gamma _{2}$ (in both cases $\tilde{\gamma}$ tends to
zero), the solutions (\ref{fund})-\ref{subharmonic} reduce to those obtained
in previous works (see e.g. \cite{PerezArj05}) for the fully dispersive
cavity. Physically a mode with large detuning or losses is strongly damped
in the cavity, so in practice it can be neglected from the very beginning.

The parametric instability can be either supercritical or subcritical,
resulting in monostable or bistable subharmonic respectively. This fact,
different to the bistability of the fundamental mode, is also present in the
fully dispersive cavity, as shown in \cite{Ostrovsky78,SanchezMor04}.

Finally we note that, in order to observe the bistable regime of the
fundamental mode, it must occur with precedence to the parametric
instability. Combining Eqs.(\ref{bistable}) and (\ref{saturatefund}), we
find that the fundamental mode presents bistability whenever%
\begin{equation*}
0<\frac{2}{3\gamma }\left( \triangle _{0}\triangle _{2}-1\right) -\triangle
_{1}^{2}<1.
\end{equation*}

\section{Experiment}

\subsection{Description}

The resonator consists in two piezoceramic discs ($\rho=7.70\times10^{3}$kg/m%
$^{3}$, $c=$m/s) with radius 1,5 cm, and thickness 1mm and 2 mm
(corresponding to resonance frequencies around 2 MHz and 1MHz,
respectively), mounted in a Plexiglas tank containing distilled and degassed
water ($\rho=1.00\times10^{3}$kg/m$^{3}$, $c=1480$m/s at $T=20^{o}C$). Both
sides are located at a variable distance L, and its parallelism can be
carefully adjusted to get a high\,-\,Q interferometer. One of the
piezoceramics, that with resonance frequency around 2 MHz is driven by the
signal provided by a function generator (Agilent 33220) and a broadband RF
power amplifier ENI 240L. The experimental setup is completed by a needle
hydrophone (TNUA200 NTR Systems) to measure the intracavity pressure field.
The Fourier transform of the acquired signal allows to determine the
spectral content of the resonator, and from it to quantify the amplitude of
any frequency component.

In this way, by changing the amplitude of the driving source, we are able to
follow the bifurcation diagram of the resonator for a given set of
parameters. Although the pump value and the decay rates $\gamma _{i}$ can be
unambiguously determined (the latter by measuring the line width of the
cavity modes), the detuning parameters $\delta _{i}$ are difficult to
control, and in general vary from one set of measurements to the next. The
reason is the dependence of the cavity resonances with the temperature of
the medium, which changes with time due to the external sources (heating of
the driving transducer or ambient temperature variations). However, we have
monitored the temperature variations to ensure that the detunings did not
changed appreciably withing one set of measurements, so we could apply the
theory of the previous section to the description of this problem.

\subsection{Bifurcation diagrams}

In the series of experiments we have obtained different bifurcation
diagrams. They differ in the spectral content (frequency of the subharmonic
pairs, weight of the higher harmonics) corresponding to different sets of
detunings, but also in the behavior of the individual amplitudes. We
illustrate two of such cases in Figs. 3 and 4.

Figure 3 was obtained for a driving frequency $f_{0}=1.991$ MHz and $L=3.5$
cm. In this case a nearly-degenerated subharmonic pair, with frequencies $%
f_{1}=1.018$ MHz and $f_{2}=0.974$ MHz, is excited. As shown in Fig. 3(a),
corresponding to a driving amplitude of 25 V, the pair is highly asymmetric,
the amplitude of one of the components, $f_{2}$, being negligible with
respect to $f_{1}$. The experimental bifurcation diagram is shown in
Fig.3(b) with symbols. The first evident feature of the diagram is the
non-linear growth of the fundamental mode amplitude as the driving amplitude 
$V_{in}$ is increased, denoting a non-trivial influence of the second
harmonic. At $V_{in}=24$ V, the threshold of parametric generation is
reached, and the amplitude of the subharmonics begins to grow. Note that in
Fig.3(b) this amplitude is magnified five times for a better clarity.

The solid lines represent the analytical stationary solutions given by Eqs.(%
\ref{fund}), (\ref{subharmonic}) and (\ref{saturatefund}). The theoretical
amplitude and pump values have been scaled by respective numerical factors
taking into account the efficiency of the transducer, the sensitivity of the
hydrophone, and the additional normalizations leading to Eqs.(\ref{adimmodel}%
). We used the medium parameters $\rho =1.00\times 10^{3}$ kg/m$^{3}$, $%
c=1480$ m/s and $\sigma =0.875$. The decay rates were obtained by measuring
the line width of the modes, resulting $\gamma _{j}\sim 5000$ s$^{-1}$.
Finally, the detunings have been chosen to get the best fit to the
experimental data. In Fig.3(b) we used $\triangle _{0}=-2.35,\,\triangle
_{1}=3.25$ and $\triangle _{2}=-4$.

For a slightly different driving frequency $f_{0}=1.879$ MHz, we observed a
different scenario, shown in Fig. 4. In this case a pair frequencies $%
f_{1}=1.242$ MHz and $f_{2}=0.638$ MHz is excited above threshold. The pair
is also clearly asymmetric in the amplitudes, as shown in Fig.4(a)
corresponding to a driving amplitude of 8 V. Different from the previous
case a linear growth of the fundamental mode below threshold is observed in
Fig.4(b), saturating at a constant value above threshold (7 V in this case)
. As discussed in the previous section, this situation is typical of a
highly detuned second harmonic, where the fully dispersive model represents
a good approach. The theoretical bifurcation diagrams in Fig.4(b) were
obtained for the set of detunings $\triangle _{0}=1.35,\,\triangle _{1}=0.6$
and $\triangle _{2}=10$, and show a good agreement with the measured values.

\subsection{Self-modulation}

Further increasing the pump above the threshold the field components often
develope low frequency side bands, as shown in Fig.5. The parameters are
those corresponding to Fig.4, and the pump parameter is nearly twice the
parametric threshold value. In the temporal domain, the presence of the side
bands implies a low frequency modulation of the field amplitudes. This
phenomenon, also demonstrated in solids \cite{Solodov04} is usually called
self-modulation since the source amplitude is kept constant. Self-modulation
can result from different mechanisms, as dicussed in \cite{Fillinger06}.

In \cite{SanchezMor06} we considered the stability of the solutions of Eqs.(%
\ref{subharmonic}) and (\ref{saturatefund}), identifying the appearance the
self-modulation with the Hopf bifurcation that the amplitudes undergo under
certain conditions. The frequency of the slow modulation predicted by the
theory was of the order of several kilohertz, for typical operation
conditions. This is in agreement with the observed side band frequency, $%
\triangle f=34$ kHz, in Fig. 5. A careful analysis of the spectrum allows to
identify these new frequencies as the result of the mixing (sum and
difference, $f_{i}\pm f_{j}$) between the fundamental and the subharmonics
components, with the second harmonics 2$f_{i}$, owing to the quadratic
nonlinearity of the medium. Some of these frequencies are correspondingly
labeled in Fig. 5. The experiment also shows that the appearance of the low
side band frequencies occurs once th pump reaches a critical value. This
behaviour of the spectrum is chracteristic of a secondary instability of the
Hopf type, in agreement with the analysis performed in \cite{SanchezMor06}.

\section{Conclusions}

The models proposed previously to describe the parametric sound generation
in acoustic resonators can not account for a number of effects observed in
the experiment. The first reasonable approach to this problem is to consider
the role played by the first higher harmonics, whose magnitude is strongly
dependent of the distance to the cavity resonances. The proposed model,
which is a simple extension of previous (dispersive) models, considers the
coupling of the fundamental field to its second harmonic. We demonstrate
that the bifurcation diagrams predicted by the theory and observed in the
experiment can be qualitatively different under different parameters, and
are very sensible to the experimental conditions (mainly the driving
frequency and the temperature, both affecting the detuning values). Such
model also predicts the emergence of secondary instabilities leading to time
self-modulated solutions, in agreement with the experimental observation at
drivings beyond the threshold. We expect that, at even higher drivings, the
dynamics of the system be chaotic, as the theory predicts. Experimental work
in the direction is in progress. 

\section*{Acknowledgments}

The work has been financially supported by the MEC of the Spanish
Government, under the project FIS2005-0793-C03-02.

\paragraph{Figure Captions}

Figure 1. Schematic representation of the field and cavity spectra, in the
cases of (a) ideal (loseless) cavity, (b) a real cavity with finite
impedances.

Figure 2. Development of bistability of the fundamental mode as the second
harmonic detuning is varied. Parameters are $\gamma =1,$$\triangle _{0}=-2,$%
\ and $\triangle _{2}=1$ (a), $\triangle _{2}=0$ (b)$,$ and $\triangle
_{2}=-1$ (c)$.$

Figure 3. Experimental bifurcation diagram (symbols) obtained for a driving
frequency $f_{0}=1.991$ MHz. Solid and dashed lines correspond to the
fundamental and subharmonic amplitudes, as obtained from Eqs. (5) and (8)
for $\triangle _{0}=-2.35,\,\triangle _{1}=3.25$ and $\triangle _{2}=-4$.

Figure 4. Experimental bifurcation diagram (symbols) obtained for a driving
frequency $f_{0}=1.879$ MHz. Solid and dashed lines correspond to the
fundamental and subharmonic amplitudes, as obtained from Eqs. (5) and (8)
for $\triangle _{0}=1.35,\,\triangle _{1}=0.6$ and $\triangle _{2}=10$.

Figure 5. Development of low-frequency side bands for the experimental
conditions in Fig. 4, for a driving amplitude $V_{in}=17$V.

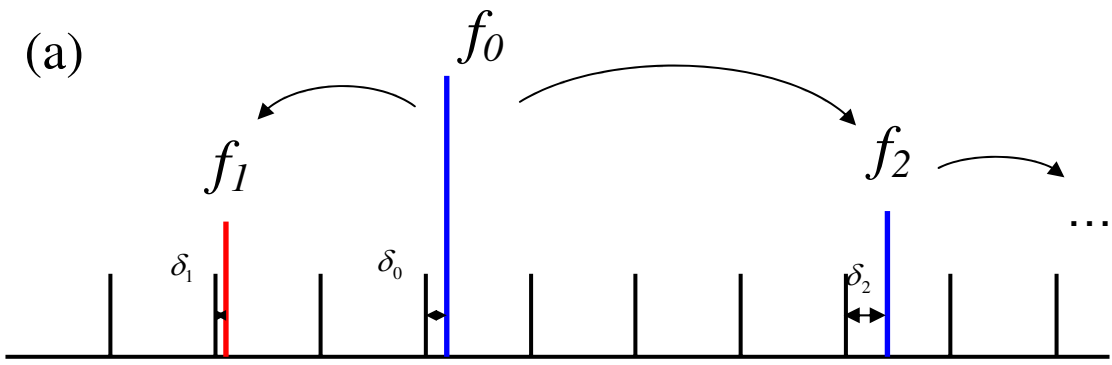
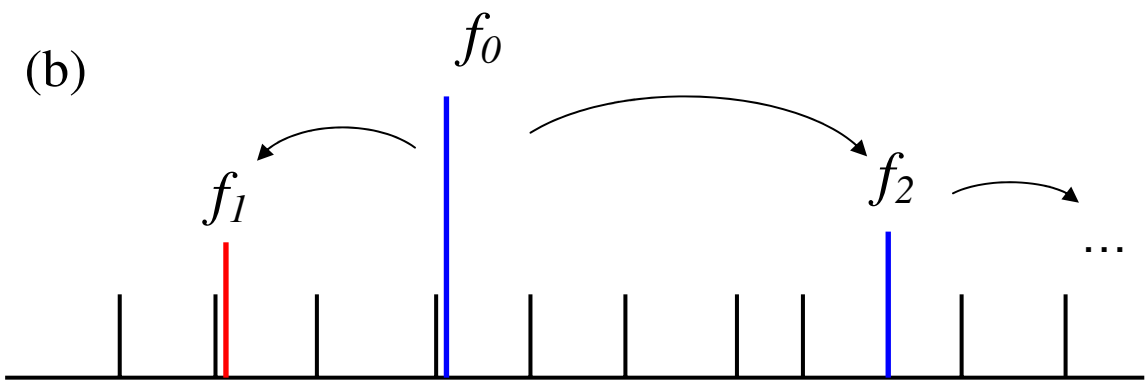
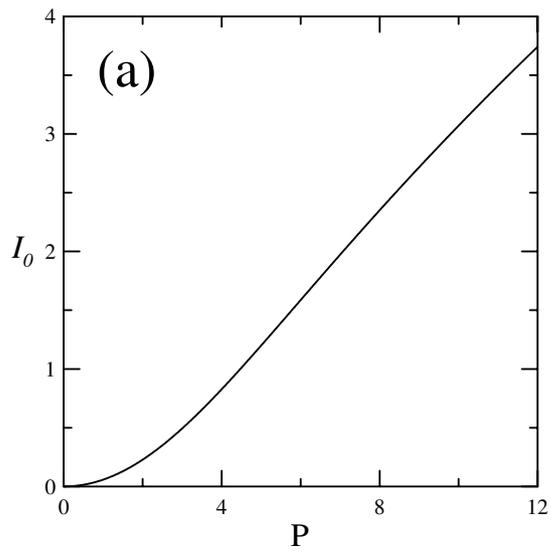
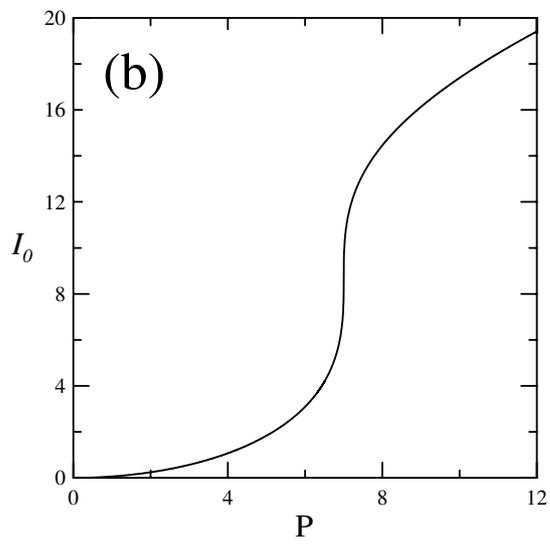
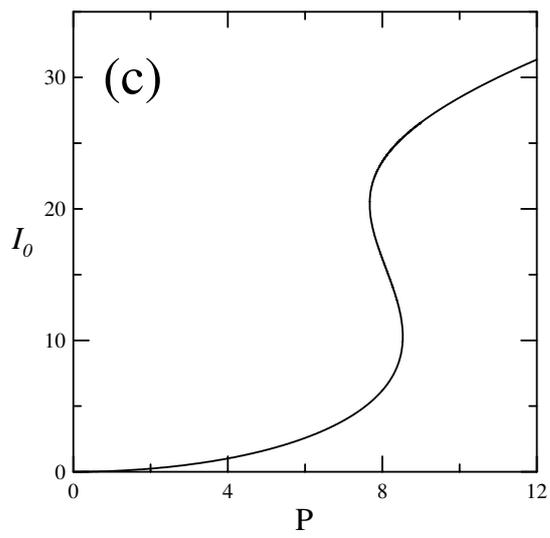
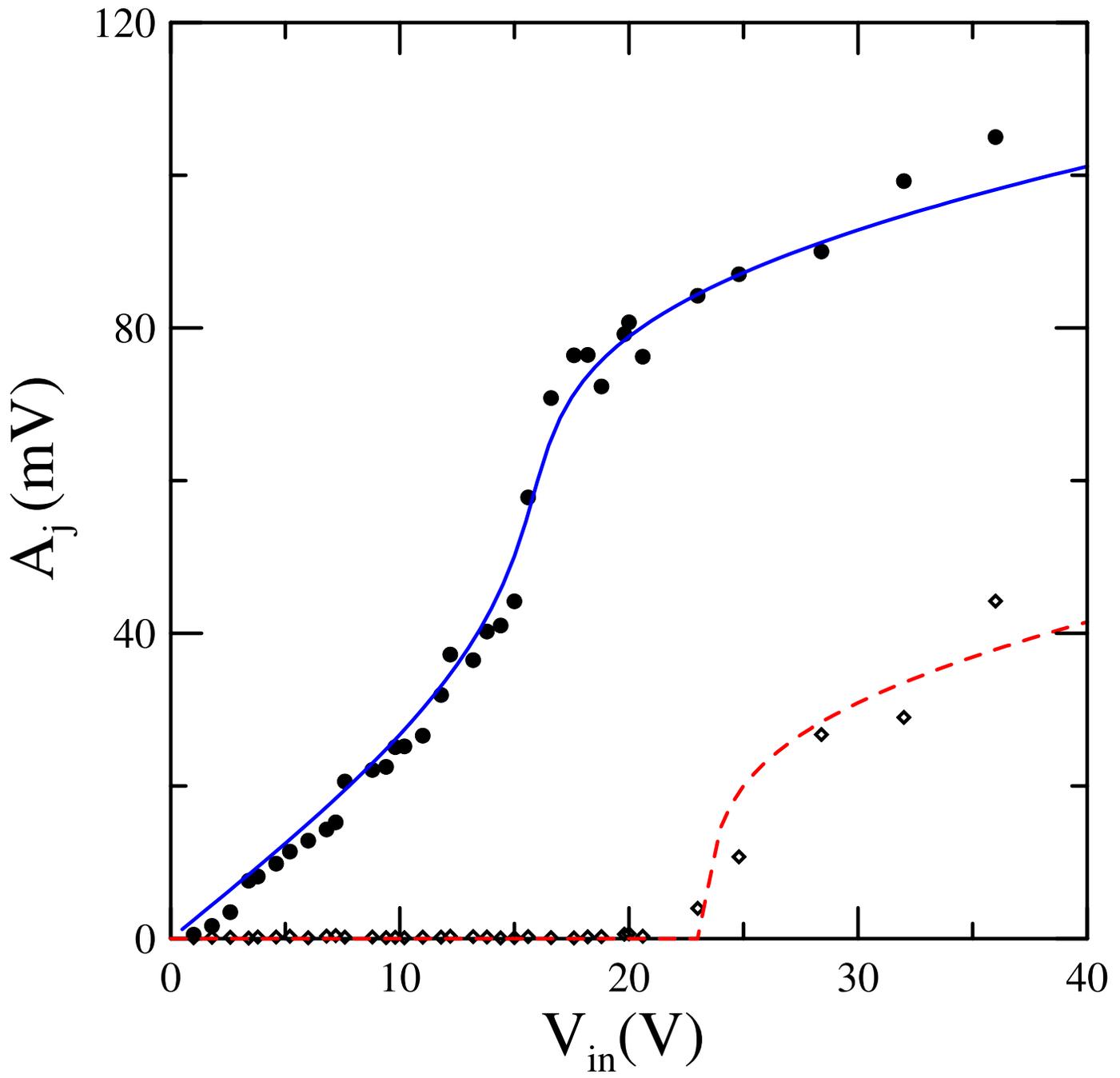
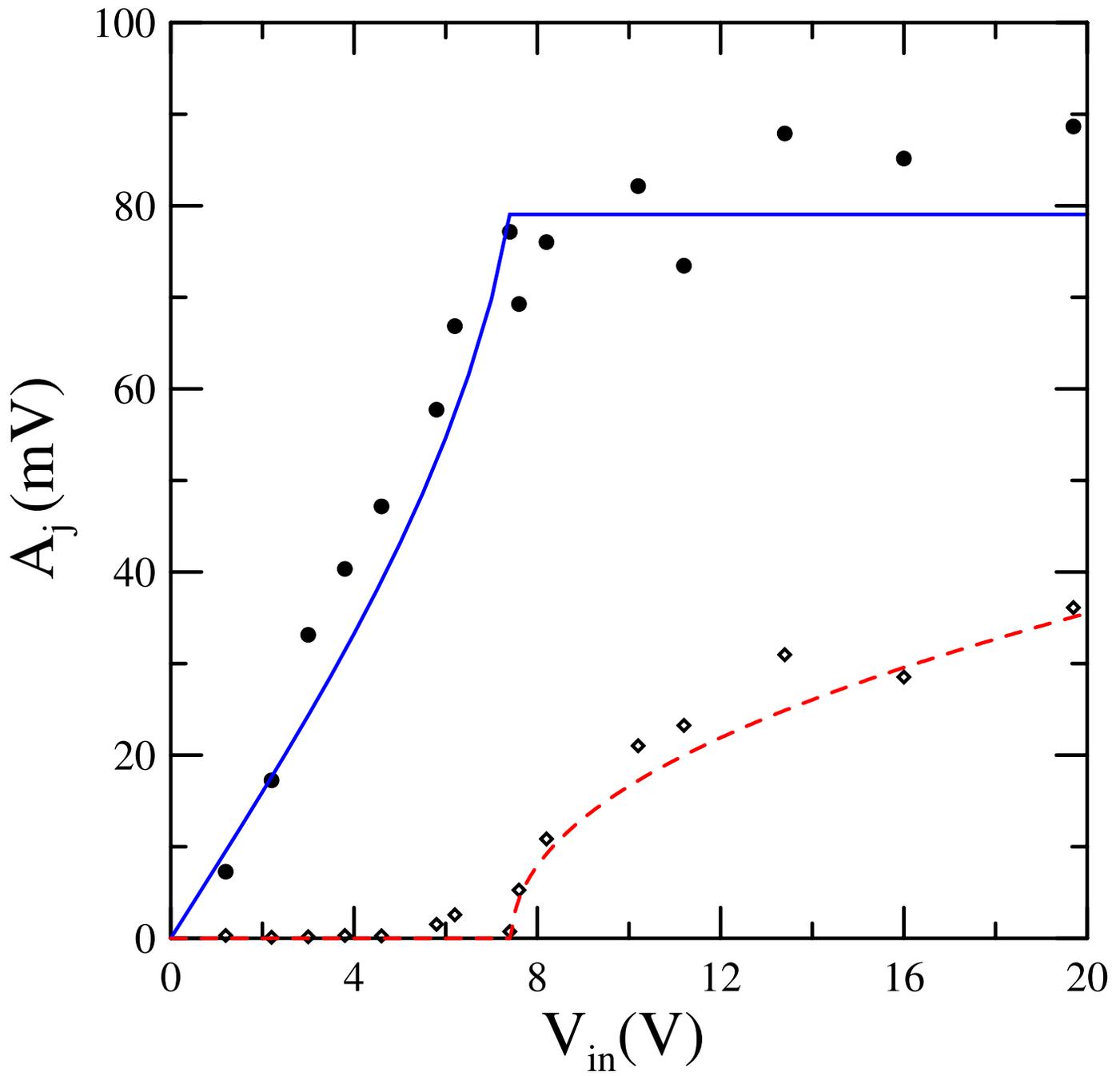
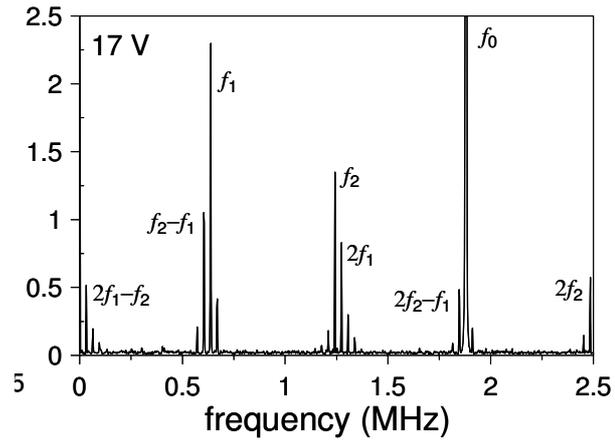
\end{document}